\documentclass[sigconf,natbib=true]{acmart}

\usepackage{tikz}
\usetikzlibrary{arrows.meta,positioning,shapes.geometric}
\usepackage{booktabs}
\usepackage{xurl}
\usepackage{subcaption}
\usepackage{xspace}

\graphicspath{{./}}

\newcommand{\skg}{\textsc{SKG}\xspace}
\newcommand{\cci}{\textsc{CCI}\xspace}
\newcommand{\dataset}{$\mathsf{DBLP}_{500K}$\xspace}

\setcopyright{none}
\acmConference[]{}{}{}
\acmYear{}
\acmDOI{}
\acmISBN{}
\begin{document}

\title[Trust-Aware Citation Cartel Ranking in Scholarly KGs]{Trust-Aware Citation Cartel Ranking in Scholarly Knowledge Graphs}

\author{Pratyush Gupta}
\email{pratyush22375@iiitd.ac.in}
\affiliation{%
  \institution{IIIT Delhi, India}
  \country{}
}

\author{Vikranth Udandarao}
\email{vikranth22570@iiitd.ac.in}
\affiliation{%
  \institution{IIIT Delhi, India}
  \country{}
}

\author{Syam Sai Santosh Bandi}
\email{syam22528@iiitd.ac.in}
\affiliation{%
  \institution{IIIT Delhi, India}
  \country{}
}

\renewcommand{\shortauthors}{Gupta, Udandarao, and Bandi}

\begin{abstract}
Citation-based systems usually treat each citation as an equal signal of scholarly influence, although citations can express very different relationships: direct method use, result comparison, broad background, or weak ceremonial acknowledgement. This distinction is crucial for citation-cartel analysis because dense internal citation alone is not suspicious; legitimate research communities are also densely connected. We present a trust-aware pipeline that combines citation graph structure with semantic citation intent to rank suspicious paper-level communities for audit. On a DBLP-derived graph with 500,000 papers and 4.87M citation edges, we use an LLM teacher to label 205,897 citation pairs, train a SciBERT student, and scale citation-intent typing to 2.04M unique graph edges. We then compute a Composite Cartel Index (\cci) that integrates internal density, citation inflation, reciprocity, semantic superficiality, degree assortativity, and trust-weighted PageRank shift. The highest-ranked community contains 1,079 papers and 8,603 internal citations, with 254.3x more internal citations than expected and 64.2\% of them superficial. Comparisons against density-only, inflation-only, semantic-only, and random baselines show that \cci cannot be reduced to a single heuristic. Edge excision validation further shows that \cci-selected communities behave differently from matched random removals. The result is a reproducible, curator-facing ranking framework for prioritising communities that warrant closer inspection.
\end{abstract}

\ccsdesc[500]{Computing methodologies~Anomaly detection}
\ccsdesc[300]{Theory of computation~Graph algorithms analysis}
\ccsdesc[300]{Applied computing~Bibliographic studies}

\keywords{scholarly knowledge graphs, citation cartels, citation intent, SciBERT, graph anomaly detection, PageRank}

\maketitle
\thispagestyle{empty} 
\pagestyle{empty}     

\section{Introduction and Related Work}
\label{sec:intro}

Scholarly knowledge graphs (\skg{}s) such as Semantic Scholar, OpenAlex, OpenCitations, and Microsoft Academic Graph encode papers and citations as large directed graphs used for search, recommendation, bibliometrics, and expert discovery~\cite{kinney2023s2ag,priem2022openalex,peroni2020opencitations,wang2020mag}. However, citation edges are semantically heterogeneous: a citation may indicate method reuse, result comparison, support, criticism, background context, or a ceremonial mention~\cite{teufel2006annotation,jurgens2018measuring,cohan2019scicite}. This limits the common assumption that citations provide uniform evidence of influence. Related work in scientometrics has similarly argued that citation impact is multidimensional and field-dependent, motivating indicators that go beyond raw citation counts or unweighted citation links~\cite{bu2019multidimensional,leydesdorff2016impact,moed2010snip}.

This ambiguity is central to identifying suspicious citation communities. Prior work has studied coercive and coordinated citation behavior at journal and author levels~\cite{wilhite2012coercive,fister2016cartels,kojaku2021detecting}, including null-model approaches that compare within-group citation exchange against expectations accounting for journal size and scientific communities. Graph anomaly detection methods identify unusual structures using density, reciprocity, degree patterns, or community structure \cite{akoglu2015graph,ma2021gad}. Yet paper-level ranking in \skg{}s remains difficult because legitimate research areas are also modular and internally dense~\cite{newman2006modularity,blondel2008louvain}. Topology-only methods can therefore conflate suspicious reinforcement with ordinary scholarly communities.

Citation-intent classification provides a complementary signal. Prior work has classified citation functions from local citation contexts and scientific-paper structure~\cite{teufel2006annotation,jurgens2018measuring,cohan2019scicite}, and recent evidence shows that filtering citation networks by intent can substantially change centrality-based paper rankings~\cite{bezerra2025citationintent}. However, this line of work has mostly treated citation intent as an edge-level NLP task or as a post-hoc analysis of network centrality, rather than using it to prioritize suspicious citation communities for audit.

We frame the problem as \emph{trust-aware ranking}: graph structure identifies where citation behavior is unusual, while citation semantics helps distinguish substantive influence from shallow reinforcement. We build a teacher--student semantic typing pipeline that scales LLM-labeled citation intents to million-edge SciBERT inference, and introduce a Composite Cartel Index (\cci) that combines structural inflation, reciprocity, semantic superficiality, assortativity, and trust-weighted PageRank shift. Our goal is not to declare misconduct, but to produce a reproducible ranked audit queue for \skg{} curators.

\vspace{3pt}
\noindent\textbf{Research Questions:} \textbf{RQ1:} Can LLM-generated citation-intent labels be distilled into a SciBERT student for million-edge semantic typing? \textbf{RQ2:} Do semantic citation signals improve suspicious-community ranking beyond topology-only baselines? \textbf{RQ3:} Do highly ranked communities exhibit graph-level behavior that differs from matched random edge removals?
\vspace{3pt}

\vspace{3pt}
\noindent\textbf{Contributions:} This work makes three contributions. First, we introduce a teacher--student semantic citation-typing pipeline that uses 205,897 LLM-labeled citation pairs to train a SciBERT student and type 2,043,874 unique citation edges. Second, we propose \cci, a community-level ranking score that combines structural anomaly signals with semantic superficiality and trust-weighted PageRank shift. Third, we validate the ranking with direct baselines, leave-one-feature ablations, multi-seed stability checks, and an edge-excision experiment comparing \cci-selected communities against matched random removals.
\vspace{3pt}

\begin{figure*}[t]
\centering
\includegraphics[width=0.90\textwidth]{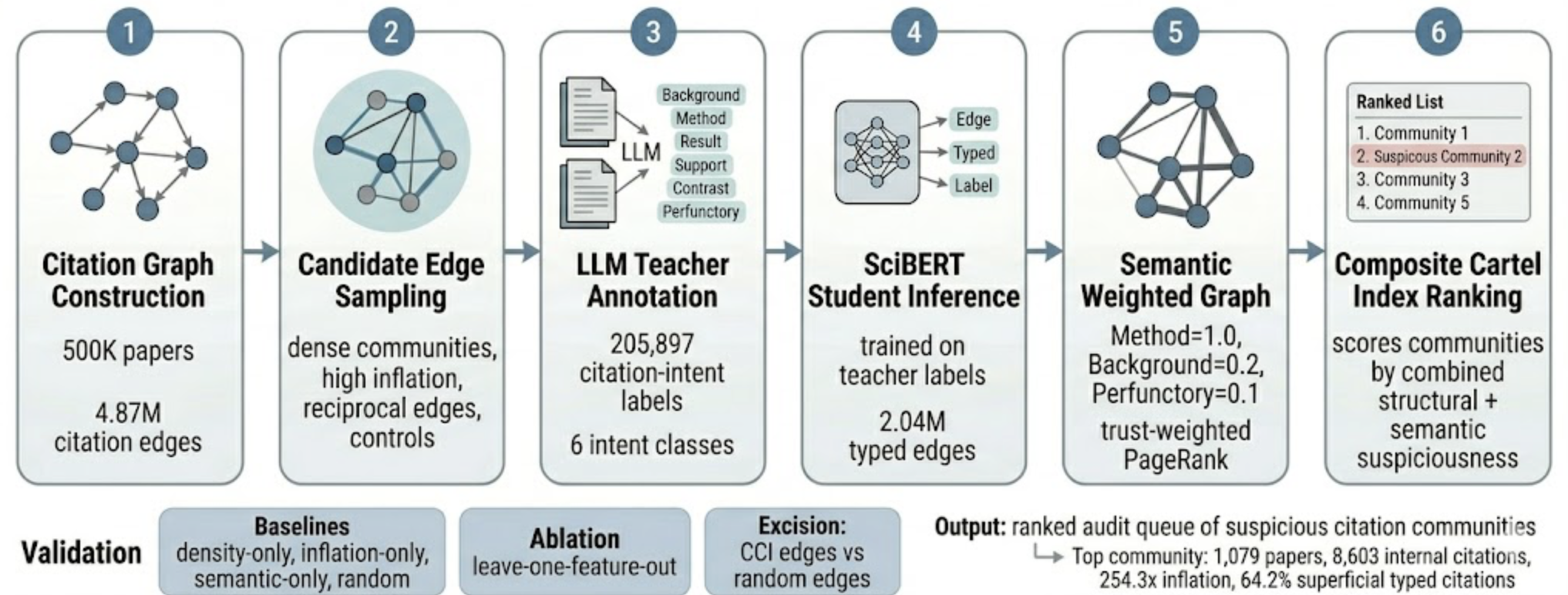}
\caption{End-to-end trust-aware citation-cartel ranking pipeline.}
\Description{A six-stage pipeline from DBLP graph construction to candidate sampling, teacher labels, SciBERT inference, semantic graph weighting, CCI ranking, and validation.}
\label{fig:pipeline}
\end{figure*}

\section{Methodology}
\subsection{Graph Topology and Profiling}
\begin{figure}[t]
\centering
\includegraphics[width=0.88\linewidth]{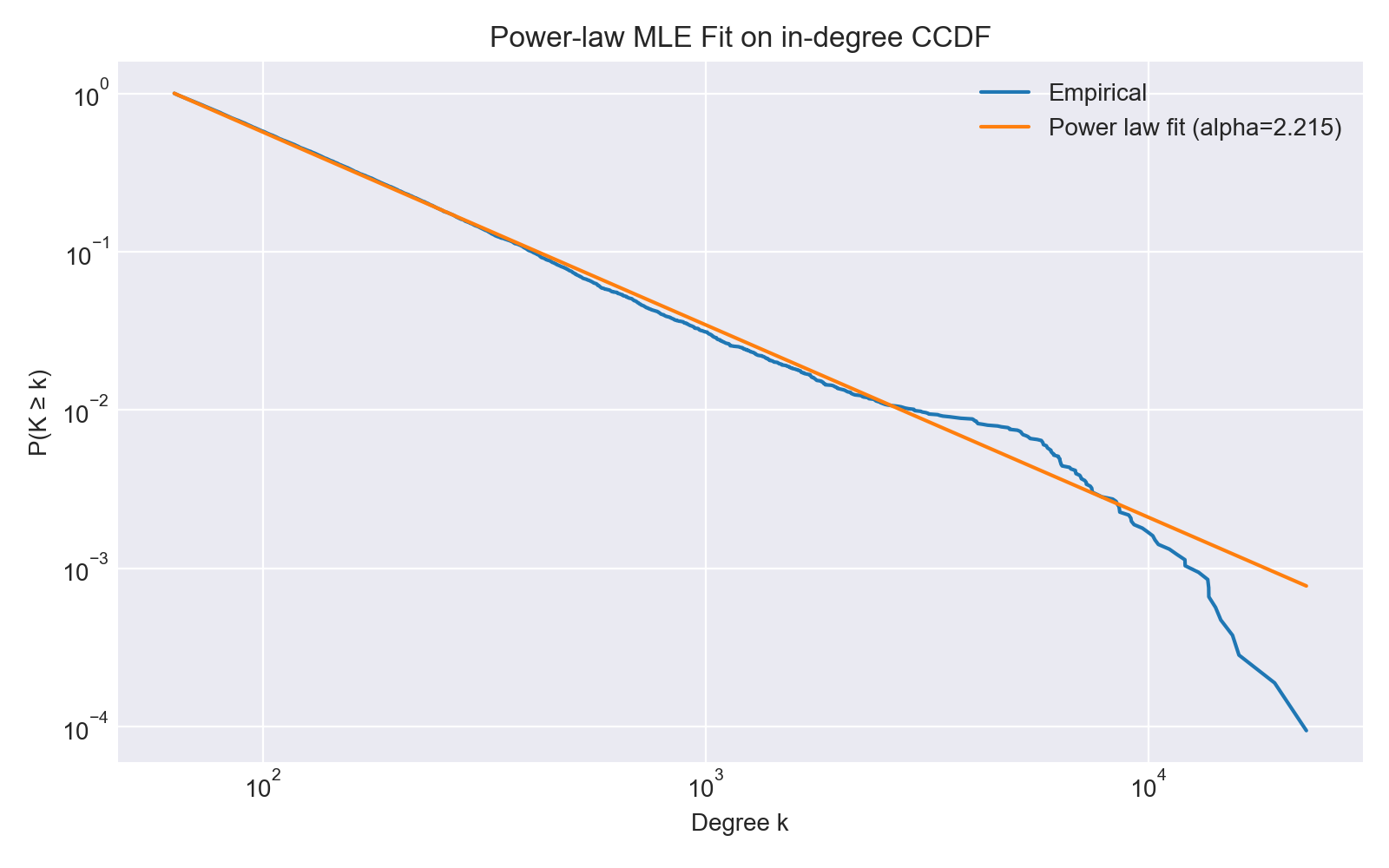}
\caption{In-degree CCDF with power-law fit ($\alpha \approx 2.215$).}
\Description{A log-log power-law CCDF fit for in-degree tracking scale-free architecture.}
\label{fig:powerlaw}
\end{figure}

\begin{figure}[t]
\centering
\includegraphics[width=0.88\linewidth]{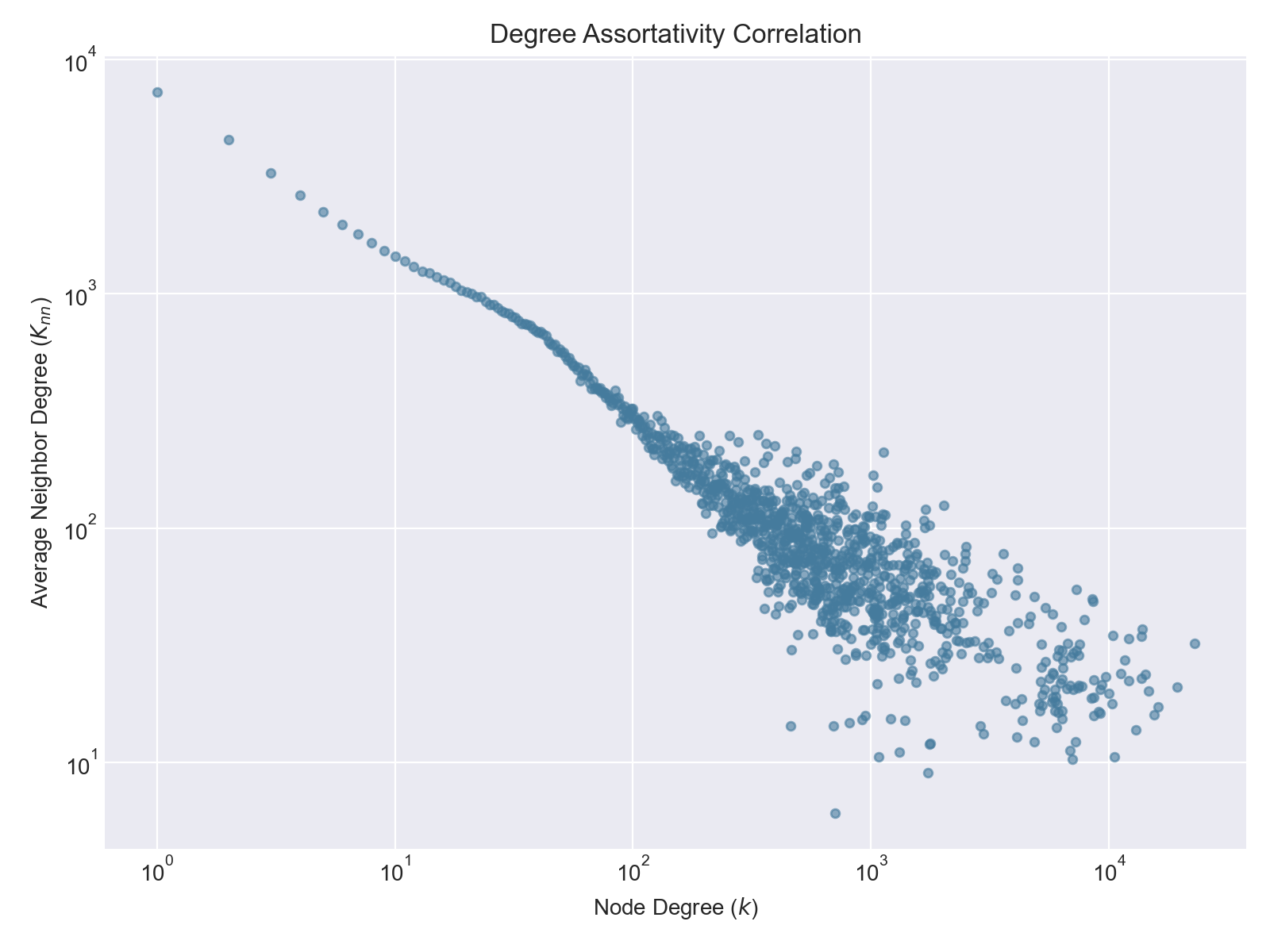}
\caption{Average neighbor degree $K_{nn}(k)$ showing disassortative mixing.}
\Description{A scatter plot of average neighbor degree vs. node degree showing a clear decreasing trend.}
\label{fig:knn}
\end{figure}

Before introducing the semantic typing pipeline, we characterize the topological structure of \dataset to motivate the design of a multi-signal anomaly score. The citation graph exhibits canonical scale-free properties~\cite{barabasi1999emergence}: the in-degree distribution follows a power law with exponent $\alpha \approx 2.215$ (Figure~\ref{fig:powerlaw}), and the mean neighbor degree decays with node degree, confirming disassortative mixing typical of citation networks~\cite{newman2001scientific} (Figure~\ref{fig:knn}). The subgraph is constructed via topological snowball sampling seeded from the 200 most-cited papers, producing a connected graph with 62.1\% of nodes having both in- and out-edges and no isolates.

The network's bow-tie decomposition reveals that 77.5\% of nodes belong to the IN-component (papers that cite into the giant strongly connected core but are not reachable from it), while only 2.6\% form the giant SCC core itself. This highly asymmetric structure means that citation flow is predominantly unidirectional: most papers cite older work without being cited back, and reciprocal citation patterns are structurally rare at the global level. When reciprocity does appear locally at elevated rates, it becomes a meaningful anomaly signal.

Robustness analysis further distinguishes the graph from random and preferential-attachment null models: targeted removal of the top-degree nodes collapses the giant component after removing only $\sim$5\% of nodes, while random failures have a gradual effect~\cite{albert2000error,callaway2000network}. This extreme hub vulnerability implies that locally dense communities whose members are \emph{not} global hubs represent a distinct structural motif, precisely the kind of pattern that citation cartels would create.

These profiling results inform three design choices in the Composite Cartel Index: (1) degree assortativity is included because the graph is globally disassortative, making locally assortative communities anomalous; (2) internal reciprocity is meaningful because global reciprocity is low; and (3) citation inflation is normalized against a degree-product baseline to distinguish genuinely excessive internal citation from the expected density of scale-free clusters.

\subsection{Data and Semantic Edge Typing}
We construct \dataset, a closed-world DBLP-derived citation graph with 500,000 papers and 4,871,544 citation edges. Edges whose cited paper is outside the sampled graph are excluded, yielding an edge-node ratio of 9.74. The closed-world construction is important: every scored citation can be attached to a paper, community, PageRank value, and local graph context. We then sample both structurally informative and control edges for semantic annotation: random edges, internal dense-community edges, high-inflation community edges, reciprocal edges, low-similarity edges, and edges touching candidate communities. This produces labels for the regions most relevant to cartel ranking while retaining background controls for comparison.

The LLM teacher assigns one of six citation-intent labels: \emph{Background}, \emph{Method}, \emph{Result/Comparison}, \emph{Support}, \emph{Contrast/Criticism}, and \emph{Perfunctory/Ceremonial}. The final teacher set contains 205,897 parsed labels, all six classes, and zero parse failures in the final batches. We train \texttt{allenai/scibert\_scivocab\_uncased}~\cite{beltagy2019scibert} with 164,717 training, 20,590 validation, and 20,590 test examples. The held-out test accuracy is 0.775, macro-F1 is 0.574, and weighted-F1 is 0.775. Rare labels remain difficult, but the model recovers cartel-relevant minority classes non-trivially: \emph{Contrast/Criticism} F1 is 0.630 and \emph{Perfunctory/Ceremonial} F1 is 0.473.

\begin{figure}[t]
\centering
\begin{subfigure}{\columnwidth}
  \centering
  \scriptsize
  \setlength{\tabcolsep}{3pt}
  \begin{tabular}{l|rrrrrr}
    \toprule
    & \rotatebox{55}{Backgr.} & \rotatebox{55}{Method} & \rotatebox{55}{Res./Comp.} & \rotatebox{55}{Support} & \rotatebox{55}{Contr./Crit.} & \rotatebox{55}{Perf./Cer.} \\
    \midrule
    Background       & \textbf{9849} & 1764 & 359 & 105 & 10  & 96  \\
    Method           & 1333 & \textbf{5573} & 127 & 28  & 0   & 28  \\
    Result/Comp.     & 333  & 154  & \textbf{298} & 8   & 1   & 0   \\
    Support          & 111  & 30   & 2   & \textbf{88}  & 0   & 1   \\
    Contr./Crit.     & 8    & 4    & 2   & 2   & \textbf{23}  & 0   \\
    Perf./Cer.       & 91   & 44   & 0   & 1   & 0   & \textbf{117} \\
    \bottomrule
  \end{tabular}
  \caption{Held-out confusion matrix (20,590 test edges).}
  \label{fig:confusion}
\end{subfigure}

\vspace{0.5em}

\begin{subfigure}{\columnwidth}
  \centering
  \scriptsize
  \begin{tabular}{lrr}
    \toprule
    Predicted intent label & Edge count & Share (\%) \\
    \midrule
    Background             & 1,169,212  & 57.2 \\
    Method                 &   749,608  & 36.7 \\
    Result/Comparison      &    80,794  &  4.0 \\
    Perfunctory/Ceremonial &    21,350  &  1.0 \\
    Support                &    19,738  &  1.0 \\
    Contrast/Criticism     &     3,172  &  0.2 \\
    \midrule
    \textbf{Total}         & \textbf{2,043,874} & \textbf{100.0} \\
    \bottomrule
  \end{tabular}
  \caption{Intent distribution over 2.04M typed edges.}
  \label{fig:intents}
\end{subfigure}
\caption{SciBERT citation-intent outputs.}
\Description{Two tables: a confusion matrix for six citation-intent labels, and an intent distribution table showing Background and Method dominate the typed edge set.}
\end{figure}

The model quality is sufficient for the role it plays in this paper: community-level semantic scoring rather than final edge-level adjudication. Most errors occur between neighboring rhetorical functions, especially Background and Method, which is expected because titles and abstracts often encode broad topical relatedness without revealing the full in-text citation context. By contrast, the model still identifies enough shallow and critical edges to construct stable aggregate features over thousands of internal citations: for example, the top-ranked community has 8,603 internal citations, so even a moderate macro-F1 classifier can provide a stable estimate of its aggregate superficiality. Figure~\ref{fig:confusion} reports the held-out confusion matrix used to audit this behavior.

We apply the trained student to a targeted large-scale inference set designed for graph analysis rather than uniform annotation: 250K initial full-graph edges, 350K structurally targeted edges, and 1.25M adaptive edges selected from communities with high structural and semantic risk. After teacher-over-student duplicate resolution, the final semantic table contains 2,043,874 unique typed edges, covering 42.0\% of the full graph. Figure~\ref{fig:intents} shows the typed-edge label distribution. This coverage is sufficient to compute community-level semantic superficiality for all major candidate communities while keeping annotation cost practical.

\subsection{Composite Cartel Index}
We map citation-intent labels to semantic trust weights as shown in Table~\ref{tab:semantic-weights}. Untyped edges receive neutral weight 1.0 for PageRank so that missing labels do not automatically penalize papers. The weighting intentionally privileges citations that encode method reuse, comparison, or evidential support and discounts citations whose observable function is broad context or ceremony. This turns the citation graph into a trust-weighted graph without deleting any edges: all structural evidence remains available, but semantically weak citations contribute less to the trust-sensitive ranking signal.

\begin{table}[t]
\centering
\caption{Citation-intent labels and semantic trust weights.}
\label{tab:semantic-weights}
\scriptsize
\begin{tabular}{lrp{0.48\columnwidth}}
\toprule
Intent label & Weight & Interpretation \\
\midrule
Method & 1.0 & Direct methodological, tool, or dataset dependence. \\
Result/Comparison & 0.7 & Comparison with related results or systems. \\
Support & 0.5 & Evidence for a claim made by the citing work. \\
Contrast/Criticism & 0.3 & Disagreement, correction, or critique. \\
Background & 0.2 & Broad but substantively relevant context. \\
Perfunctory/Ceremonial & 0.1 & Weak, generic, or ceremonial citation relation. \\
\bottomrule
\end{tabular}
\end{table}

For each Louvain community~\cite{blondel2008louvain}, we compute six features: (1) internal directed density; (2) citation inflation, the ratio between observed internal citations and degree-product expectation; (3) internal reciprocity; (4) semantic superficiality, the fraction of typed internal edges whose label is Background or Perfunctory/Ceremonial; (5) log-degree assortativity over internal edges; and (6) mean PageRank drop from unweighted to trust-weighted PageRank~\cite{page1999pagerank}. These features capture complementary failure modes: excessive local citation concentration, shallow semantic support, reciprocal reinforcement, and rank sensitivity after trust weights are applied. The \cci score is the mean of feature z-scores. We use \cci as a ranking function for audit.

\section{Results and Discussion}
Table~\ref{tab:top} reports the highest-ranked communities. The top community is compact relative to broad fields (1,079 papers) but has 8,603 internal citations, 254.3x inflation over expectation, and 64.2\% superficiality among typed internal edges, making it a strong audit candidate. Larger communities also appear in the top five, but their inflation and semantic profiles differ: some are large high-superficiality regions, while others are smaller high-inflation structures. This diversity is precisely why a multi-feature score is more useful than a single density threshold.

The results confirm that semantic typing changes the operational form of cartel detection. Suspicious communities are dense in ways that uniquely combine inflation, shallow function, and trust-rank sensitivity, making the index highly interpretable for curators.

\begin{table}[t]
\centering
\caption{Top \cci-ranked communities. Sup. is semantic superficiality over typed internal edges.}
\label{tab:top}
\scriptsize
\begin{tabular}{rrrrrrr}
\toprule
Rank & Comm. & Papers & Int. edges & Infl. & Sup. & \cci \\
\midrule
1 & 14 & 1,079 & 8,603 & 254.3 & 0.642 & 1.863 \\
2 & 15 & 22,125 & 138,436 & 29.2 & 0.800 & 0.441 \\
3 & 0 & 16,575 & 68,851 & 37.7 & 0.705 & 0.402 \\
4 & 5 & 6,960 & 66,597 & 29.4 & 0.539 & 0.373 \\
5 & 13 & 12,052 & 122,138 & 37.2 & 0.850 & 0.234 \\
\bottomrule
\end{tabular}
\end{table}

\begin{figure}[t]
\centering
\includegraphics[width=0.88\linewidth]{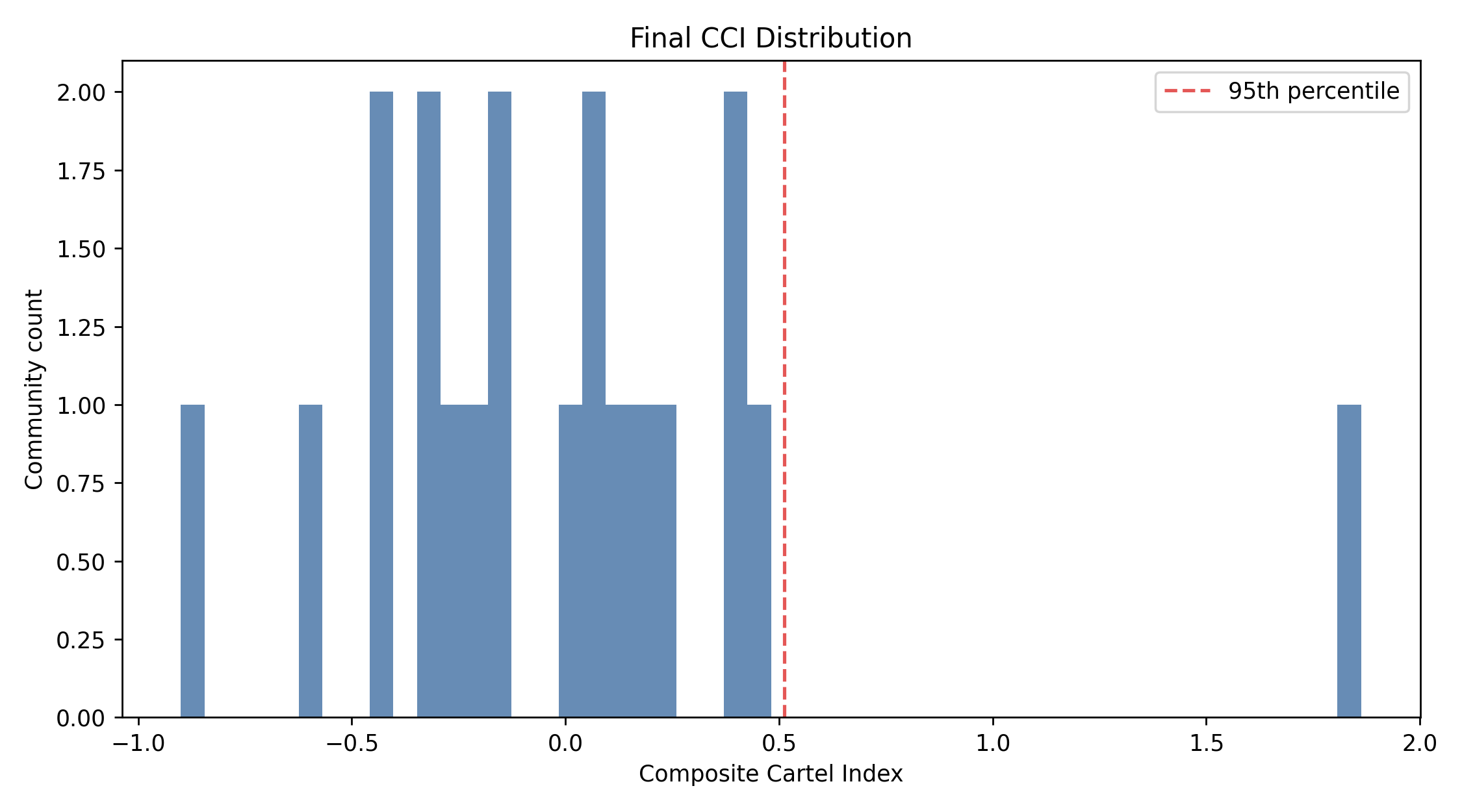}
\caption{\cci distribution across isolated candidate communities.}
\label{fig:cci_dist}
\end{figure}

\begin{figure}[t]
\centering
\includegraphics[width=0.88\linewidth]{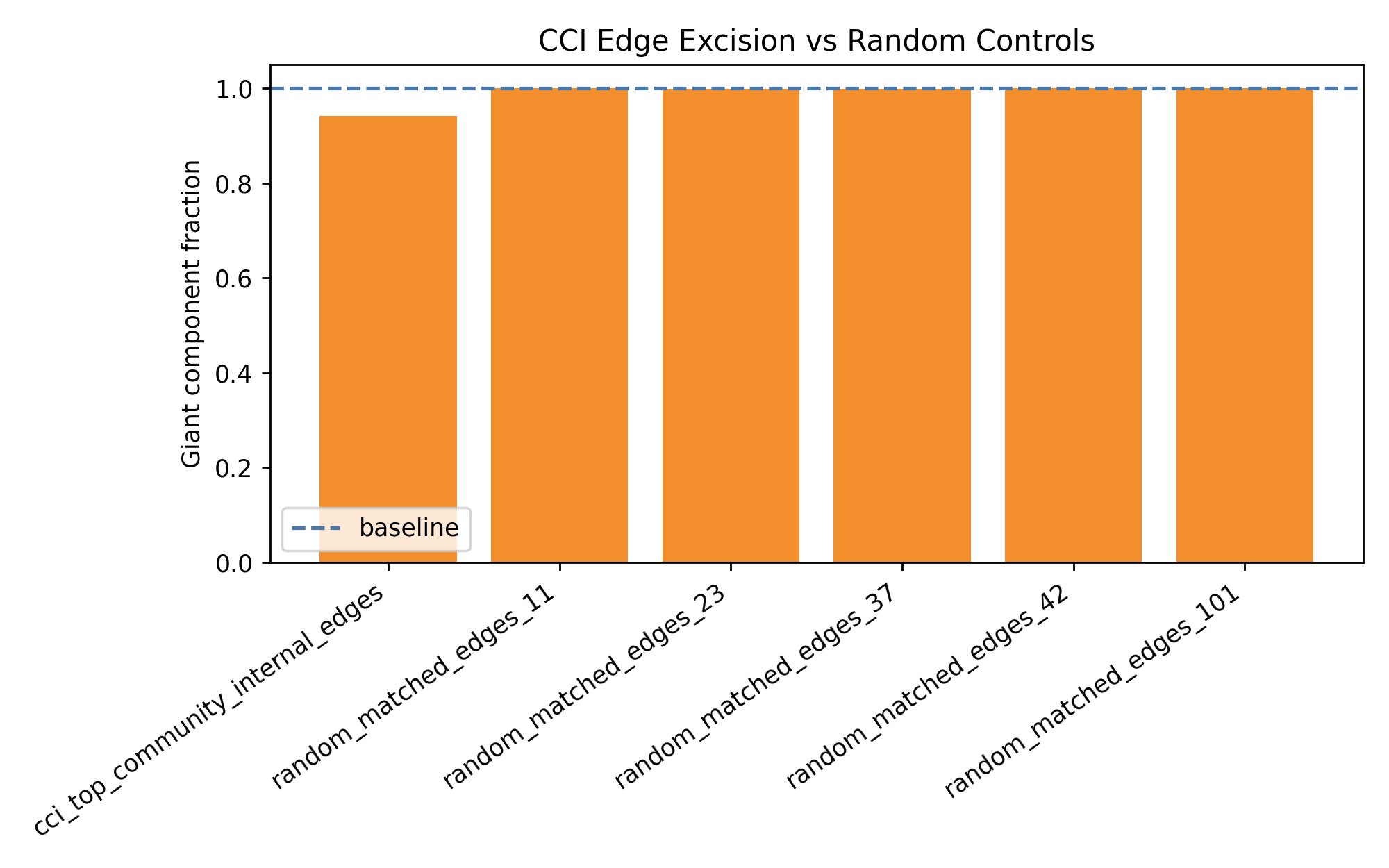}
\caption{Edge excision stress test tracking giant component stability.}
\label{fig:excision}
\end{figure}

\subsection{Baselines, Ablations, and Excision}
Table~\ref{tab:checks} summarizes the paper-facing validation checks. Density alone has only 0.513 Spearman correlation with the final \cci ranking; the structural-only variant is closer (0.875) but loses semantic information. The semantic-only baseline is also incomplete: it overlaps with 40\% of the final top five. Thus, \cci does not merely restate a known graph heuristic. The leave-one-feature ablations (Table~\ref{tab:ablation}) show the same pattern from the opposite direction: the ranking remains coherent when one feature is removed, but top-five membership changes for PageRank drop and semantic superficiality, indicating that these trust-aware signals affect the audit queue.

\begin{table}[t]
\centering
\caption{Ranking checks. P@k denotes overlap with the final \cci top-k ranking.}
\label{tab:checks}
\scriptsize
\begin{tabular}{lrrr}
\toprule
Scoring variant & Spearman & P@5 & P@10 \\
\midrule
Density only & 0.513 & 0.60 & 0.60 \\
Inflation only & 0.672 & 0.40 & 0.80 \\
Reciprocity only & 0.068 & 0.40 & 0.60 \\
Structural-only \cci & 0.875 & 0.80 & 0.70 \\
Semantic-only superficiality & 0.680 & 0.40 & 0.90 \\
Random & 0.174 & 0.20 & 0.60 \\
\bottomrule
\end{tabular}
\end{table}

\begin{table}[t]
\centering
\caption{Leave-one-feature ablation of \cci.}
\label{tab:ablation}
\scriptsize
\begin{tabular}{lrrr}
\toprule
Removed feature & Spearman & P@5 & P@10 \\
\midrule
Density & 0.998 & 0.80 & 1.00 \\
Inflation & 0.904 & 0.80 & 0.80 \\
Reciprocity & 0.884 & 0.80 & 0.90 \\
Semantic superficiality & 0.875 & 0.80 & 0.70 \\
Degree assortativity & 0.977 & 1.00 & 0.90 \\
PageRank drop & 0.886 & 0.60 & 0.90 \\
\bottomrule
\end{tabular}
\end{table}

For a graph-level stress test, we remove all 404,625 internal edges in the top five \cci communities and compare with matched random removals~\cite{albert2000error,callaway2000network,cohen2001breakdown}. The targeted removal leaves 94.1\% of nodes in the giant component, whereas random removals leave 99.95--99.96\%. This gap shows that \cci-selected edges form concentrated local structures with a distinct graph signature from random deletion, while most of the global citation backbone remains intact. Removing 405K targeted edges affects about 5.8 percentage points more nodes than equivalent random removal, confirming that these communities are locally cohesive but globally peripheral. For curators, this yields an actionable workflow backed by clear data provenance.

\section{Conclusion}
We introduced a trust-aware pipeline for paper-level citation-cartel ranking in scholarly knowledge graphs. By combining LLM-supervised citation-intent typing, SciBERT inference, weighted PageRank, and multi-feature community scoring, the method produces a reproducible audit table rather than a brittle topological alarm. On a 500K-paper DBLP graph, it surfaces compact structural-semantic outliers, quantifies how they differ from simple baselines, and produces paper-ready artifacts for curator review.

\appendix
\section{Reproducibility Artifacts}
Pipeline artifacts are saved at each stage: teacher labels, semantic edges, scores, baselines, and ablations are provided in the supplementary material archive.

\section{CCI Feature Definitions}
\cci averages six dimensions for community $C$: (1) internal density, (2) inflation (observed edges vs. expectation), (3) reciprocity (fraction of mutual edges), (4) semantic superficiality (fraction of Background/Perfunctory labels), (5) degree assortativity, and (6) PageRank drop under structural trust downweighting.

\section{Validation Protocol}
Evaluation spans three axes: baseline checks against separate graph heuristics, leave-one-feature ablations to quantify rank robustness, and an edge-excision paradigm verifying global component layout resilience across 404,625 coordinated deletions.

\newpage

\section{Generative AI Usage Disclosure}
Generative AI was used in two phases of this work. First, Azure OpenAI
GPT-4.1-mini was used as a teacher model to label 205,897 citation-intent
examples. These labels were used only as supervision for the local SciBERT
student model; the final large-scale graph inference and all graph metrics were
computed by the student model and deterministic analysis scripts. Second,
GenAI-assisted writing tools were used to proofread and improve the clarity of
the manuscript. The study design, data processing, feature engineering,
mathematical definitions, experiments, and empirical conclusions remain the
authors' work.

\end{document}